\documentclass[useAMS,usenatbib,usegraphicx]{mn2e}
\usepackage{amssymb}
\usepackage{graphicx}
\usepackage{subfigure}
\usepackage{longtable}

\newif\ifAMStwofonts

\def\kms{km\,s$^{-1}$}

\newcommand{\hi}{H\,{\sc i}}
\newcommand{\hii}{H\,{\sc ii}}
\newcommand\msun{M$_\odot$}
\newcommand\etal{et al.~}

\title{Sh2-205: I. The surrounding interstellar medium}
\author[G.A. Romero and C.E. Cappa]
       {G.A. Romero$^{1,2}$\thanks{E-mail: gisela@iar-conicet.gov.ar} and C.E. Cappa$^{1,2}$\\
      $^{1}$Instituto Argentino de Radioastronom\'{\i}a, CC 5, 1894 Villa Elisa,
       Argentina,                                             \\
      $^{2}$Facultad de Ciencias Astron\'omicas y Geof\'{\i}sicas,
       Universidad Nacional de La Plata,                      \\
       Paseo del Bosque s/n, 1900 La Plata, Argentina}

\date{Accepted 2008 February 15.  Received 2008 February 14; in original form 2007
December 18}
       
\pagerange{\pageref{firstpage}--\pageref{lastpage}}

\pubyear{2008}

\begin{document}

\maketitle

\label{firstpage}

\begin{abstract}
We investigate the distribution of the interstellar matter in the environs
of the \hii\ region \hbox{Sh2-205}, based on neutral hydrogen 21--cm line data and radio continuum images at 408 and 1420 MHz obtained from the CGPS, $^{12}$CO (J=1$\rightarrow$0) observations, high resolution {\it IRAS} data ({\it HIRES}), and
MSX data.\\
Sh2-205 can be separated in three independent optical structures: \hbox{SH\,149.25--0.0}, \hbox{SH\,148.83--0.67}, and \hbox{LBN\,148.11--0.45}. The three regions are detected both at 408 and 1420 MHz. The derived spectral indices show the thermal nature of \hbox{SH\,148.83--0.67} and LBN\,148.11--0.45.\\
LBN\,148.11--0.45 is a classical \hii~region surrounded by an \hi~shell. The associated neutral atomic and ionized masses are 65 \msun~and 70 \msun, respectively. Dust and molecular gas ($\approx$ 6 $\times$ 10$^{4}$ \msun) are found related to this ionized region. Particularly, a photodissociation region is detected at the interface between the ionized and molecular regions. If the proposed exciting star HD\,24094 were an O8--O9 type star, as suggested by its near-infrared colors, its UV photon flux would be enough to explain the ionization of the nebula. \\
The optical, radio continuum, and 21--cm line data allow us to conclude that \hbox{SH\,148.83--0.67} is an interstellar bubble powered by the energetic action of \hbox{HD\,24431}. The associated neutral atomic and ionized masses are 180 \msun~and 300 \msun, respectively. The emission of \hbox{SH\,149.25--0.0} is too faint to derive the dust and gas parameters.\\
An \hi~shell centered at \hbox{{\it (l,b)} = (149\degr 0\arcmin, --1\degr 30\arcmin)} was also identified. It correlates morphologically with molecular gas emission. The neutral atomic and molecular masses are 1600 \msun~and 2.6 $\times$ 10$^{4}$ \msun, respectively. The open cluster NGC\,1444 is the most probable responsible for shaping this \hi~structure.
\end{abstract}

\begin{keywords}
stars: individual: HD\,24431, ALS\,7793, and HD\,24094 - ism: structure - ism: bubbles
\end{keywords}

\section{Introduction}\label{intro}

O-type stars emit a large amount of UV photons capable of ionizing the
neutral atomic gas and photodissociating molecular material, creating
\hii\ regions (Osterbrock 1989) and photodissociation regions at the interface
between the ionized and the molecular gas (Hollenbach \& Tielens~1997). \hii\ regions evolve
from ultracompact to classical modifying the characteristics of their
natal environment. They appear surrounded by neutral atomic and molecular gas, which generally constitute the remains of their natal clouds. Subsequent dynamical expansion favors the formation of
dense low expanding neutral envelopes behind front shocks (Spitzer~1978) where conditions for stellar formation may be fulfilled.

Massive stars also lose mass at rates of \hbox{10$^{-6}$ -- 10$^{-7}$\,M$_\odot$\,yr$^{-1}$} with terminal velocities of \hbox{1000 - 2000 \kms}~(Prinja \etal1990; Chlebowski \& Garmany 1991; Lamers \& Leitherer
1993). Although clumpy stellar winds result in lower mass
loss
rates (Moffat \& Puls~2003), the huge amount of mechanical energy released into the
interstellar medium (ISM) through the stellar wind mechanism during the lifetime of a massive star is capable of
creating stellar wind bubbles. An outer dense neutral shell will appear
between the outer shock front and the ionization front if this last front is trapped within the expanding bubble (Weaver \etal1977; Lamers \& Cassinelli~1999).

The action of massive stars on their surrounding interstellar medium can be analyzed
using radio continuum data, which allow us to know the
characteristics
of the ionized gas. Neutral hydrogen 21--cm line emission and molecular
data allow us to investigate the distribution of the neutral atomic and molecular gas in the environs of the stars, while
interstellar dust characteristics can be studied through its infrared
 emission.

Massive stars are the main source of energy injection into the interstellar medium in the Galaxy. This causes the re-distribution of interstellar material by photodynamical and photochemical processes. To improve theoretical scenarios, comprehensive observational studies are necessary. This is possible by achieving complete multiwavelength studies of \hii\ regions and interstellar bubbles.\\
\indent To contribute to the knowdledge of the ISM and the processes that modify it, we are carrying out a systematic study of the kinematics and
energetics of the ionized and neutral gas in the environs of massive stars.  
In this paper we specially analyze the \hii\ region \hbox{Sh2-205} in the region of Camelopardalis based on radio continuum, \hi\ 21--cm line, $^{12}$CO line, and infrared data at several wavelengths.

\section{The \hii~regions and their exciting stars}\label{stars}

Figure 1 displays the VTSS and DSS-R images of a region of \hbox{$\approx$ 2\degr$\times$2\degr} centered at \hbox{{\it (l,b)} = (148\degr40\arcmin, --0\degr20\arcmin)}, showing the \hii~region Sh2-205 (Sharpless~1959) of about 120\arcmin~in size. The images reveal the presence of diffuse extended emission with two bright areas. One of them consists of an arc-like structure located at \hbox{{\it (l,b)} = (148\degr 45\arcmin, --0\degr 50\arcmin)}. The VTSS image shows that the edges of this arc are brighter than the inner region. Along with the fainter optical emission region detected at \hbox{{\it (l,b)} $\thickapprox$ (149\degr 0\arcmin, --0\degr 20\arcmin)}, the arc-like feature delineates an almost complete shell centered approximately at \hbox{{\it (l,b)} = (148\degr50\arcmin, --0\degr40\arcmin)}, of $\approx$ 60\arcmin~in diameter. From here on, this structure will be referred to as SH\,148.83--0.67.

The exciting star of this \hii~region is the O-type star HD\,24431 (Sharpless~1959), located at \hbox{{\it (l,b)} = (148\degr 50\arcmin, --0\degr 43\arcmin)},  projected close to the center of this structure. This star was classified as O9 IV-V by Hiltner \& Johnson~(1956) and as O9 III by  Walborn~(1973). Fabricius \etal(2002), based on Tycho2 data identifies this star as a double system.

 \begin{figure}
 \centering
 \label{optico}
 \includegraphics[angle=0,width=8cm]{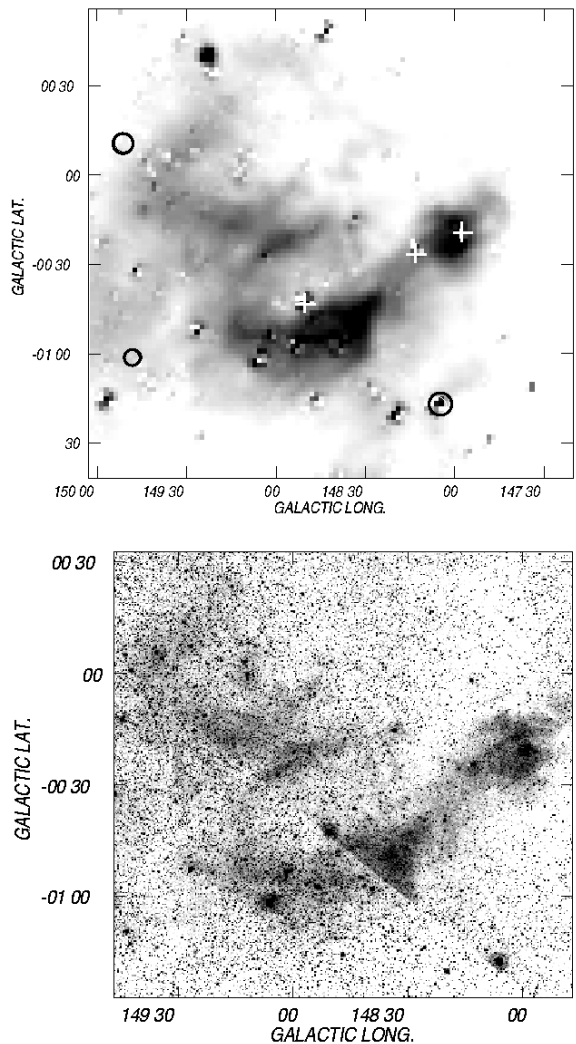}
 \caption{\small{Optical images of the Sh2-205 \hii~region. {\it Top panel}: VTSS Continuum-corrected H$\alpha$ image. The positions of the stars HD\,24431, ALS\,7793, and HD\,24094 are indicated by crosses. The locations of the open clusters NGC~1496, King~7 and NGC~1444 are marked by open circles. {\it Botton panel}: DSS-R image of the region.}}
 \end{figure}

Distance determinations for HD\,24431 span the range between 710 and 1000 pc (Zdanavi\v{c}ius \etal2001; Gies~1987; Garmany~1982; Humphreys~1978). Cruz-Gonz\'alez \etal(1974) adopted a distance of 2.28 kpc, considering this star to be a member of $\chi$ Persei. However, this open cluster is located at \hbox{{\it (l,b)} = (135\degr,--3\degr 36\arcmin)}, 14\degr~far from HD\,24431. This fact makes its relation to the cluster doubtful.

Table\,1 lists the main photometric and astrometric parameters of HD\,24431 taken from the literature. The star was related to Cam~OB1 ({\it [l,b]} = [142\degr, +2\degr] of $\approx$  1\fdg7 $\times$10\degr~in size) by Humphreys~(1978) and Garmany~(1982). Lyder~(2001) carried out an extensive analysis of the optically identified Population~I objects in the Cam~OB1 region, deriving a distance of 975~$\pm$~90 pc for the OB association. This author, however, does not consider HD\,24431 to be a member of Cam~OB1. In what follows, we adopt a distance \hbox{{\it d}~=~1.0 $\pm$~0.2 kpc} for HD\,24431, in agreement with previous determinations.

\begin{table}
 \caption{Photometric and astrometric results for HD\,24431, ALS\,7793, and HD\,24094.}
\begin{tabular}[h]{|c|c|l|c|c|c|c|}
\hline\hline
\scriptsize
~&\multicolumn{6}{c}{HD\,24431}\\
\cline{1-7}
\cline{1-7}
\it{l}&\it{b}&Sp.T.&V$_{r}$(\scriptsize{LSR})&\multicolumn{2}{c}{Distance}\\
{[\degr]}&[\degr]&             &[kms$^{-1}$]&[pc]&Ref.\\ [10pt]
148\degr50\arcmin&--0\degr43\arcmin&O9 IV-V~$^{1}$&--9.9$\pm$~2~$^{3}$ &957&4\\
~&~&O9 III~$^{2}$&~&1000&3,5\\
~&~&~&~&~2280&6\\
\hline
~&\multicolumn{6}{c}{ALS\,7793}\\
\hline
\it{l}&\it{b}&Sp.T.&m$_{v}$&&&\\
{[\degr]}&[\degr]& &[mag] &&\\ [10pt]
148\degr 17\arcmin& --0\degr 26\arcmin&B1 V~$^{1}$&11.1~$^{1}$&&&\\
\hline
~&\multicolumn{6}{c}{HD\,24094 $^{7}$}\\
\hline
\it{l}&\it{b}&Sp.T.&P&B-V&&\\
\multicolumn{1}{c}{[\degr]}&[\degr]& &[days]&[mag]&&\\[10pt]
147\degr 58\arcmin&--0\degr 20\arcmin&B8 V&1.052&0.4&&\\
\hline\hline
\end{tabular}\\
\scriptsize
Notes: 1: Hiltner \& Johnson~(1956); 2: Walborn~(1973); 3: Humphreys~(1978); 4: Zdanavi\v{c}ius \etal2001; 5: Gies~(1987); Garmany~(1982); 6: Cruz-Gonz\'alez \etal(1974); 7: Duerbeck~(1997).
\end{table}

\normalsize
The other bright extended area shown in Fig.\,1 is the conspicuous \hii~region centered at \hbox{{\it (l,b)} = (148\degr 5\arcmin, --0\degr 28\arcmin)}. This region was catalogued as LBN\,148.11--0.45 (43\arcmin$\times$25\arcmin~in size) by Lynds~(1965), although it  had originally been included in Sh2--205 by Sharpless~(1959). Two B-type stars appear projected onto this region. One of them is ALS\,7793 (\hbox{{\it [l,b]} = [148\degr 17\arcmin, --0\degr 26\arcmin])}, whose spectral type is B1 (Hiltner~1956). The star is situated onto the border of the \hii~region. The other star, HD\,24094 (CY~Cam, \hbox{{\it [l,b]} = [147\degr 58\arcmin, --0\degr 20\arcmin])}, is seen projected near the center of LBN\,148.11--0.45. Based on the Hipparchos catalogue, HD\,24094 was identified as a slowly pulsating contact binary by Duerbeck~(1997), its primary component being classified as a B8 star. The main parameters of these stars are listed in Table\,1. No information about the distances to HD\,24094 and ALS\,7793 is available. As a working hypothesis, we adopt \hbox{{\it d}~=~1.0~$\pm$~0.2~kpc}.

Radio studies of the Sh2-205 \hii~region have been performed by several authors. Blitz, Fich \& Stark~(1982) detected a molecular component at \hbox{--25 $\pm$ 1 \kms} in their CO study of \hii~regions, pointing the telescope towards the position \hbox{{\it (l,b)} = (148\degr 50\farcm4, --1\degr 14\farcm4)}. Based on radio continuum data at \hbox{1400~MHz} with an angular resolution of 10 arcmin, Felli \& Churchwell~(1972) found two radio sources. One of them, centered at \hbox{{\it (l,b)} = (147\degr 54\arcmin, --0\degr 24\arcmin)}, has a flux density of 7.4 Jy and is 30\farcm7 $\times$ 22\farcm7 in size. This source spatially coincides with LBN\,148.11--0.45. The other radio source was detected at  \hbox{{\it (l,b)} = (148\degr 42\arcmin, --0\degr 7\arcmin)}, has a flux density of 0.2 Jy, and is 14\arcmin $\times$ 10\farcm8 in size. It appears projected onto a region almost free of optical emission.

 No radial velocity information corresponding to the ionized gas in the bright regions is available. The only measurement comes from Fabry-Perot H$\alpha$ observations obtained by Fich, Dahl \& Treffers~(1990), who found a radial velocity of $\hbox{--16.8 $\pm$ 0.3 \kms}$ observing towards the position \hbox{{\it (l,b)} = (147\degr 55\farcm8, --0\degr 3\farcm46)}.

In addition to the above mentioned optical features, a shell-like structure of faint diffuse optical emission is also detected centered at \hbox{{\it (l,b)} = (149\degr 15\arcmin,  0\degr)} (from here on SH\,149.25--0.00), spread over a circle of $\sim$ 1\degr~in size. No stellar object seems to be related to this feature. This structure, together with the one probably related to HD\,24431, were catalogued as \hbox{LBN\,149.02--00.13} by Lynds~(1965).

The positions of  the open clusters NGC~1496 at \hbox{{\it(l,b)} = (149\degr 51\arcmin, +0\degr 11\arcmin)},  NGC~1444 at \hbox{{\it (l,b)} = (148\degr 6\arcmin, --1\degr 18\arcmin)}, and King~7 at \hbox{{\it (l,b)} = (149\degr 46\arcmin,  --1\degr 1\arcmin)} are indicated in the VTSS image. Their distances are 1.2 kpc (del Rio \& Huestamendia~1988), 1.19 kpc (Lindoff~1968), and 2.2 kpc (Durgapal \etal1997), respectively. Although they are seen projected close to the regions under study, their physical relation to the optical nebulae is doubtful.

\section{Data sets}

Radio continuum images at 408~MHz and 1420~MHz, and 21--cm \hi~line emission data were taken from the Canadian Galactic Plane Survey (CGPS) observed with the DRAO Synthesis Telescope, in Penticton, Canada. The CGPS is a high resolution survey of atomic hydrogen and radio continuum emission from the Galaxy. The spectral line observations are  presented as data-cubes with 272 spectral channels having a velocity resolution of 1.3 $\hbox{km s$^{-1}$}$. Table\,2 presents the most important observational parameters of the DRAO data base (Taylor \etal2003). Both the radio continuum images at 408~MHz and 1420~MHz, and \hi~line data cube were spatially smoothed to an angular resolution of $\sim$2 arcmin.

The radio continuum image at 2695~MHz, obtained using the Effelsberg 100-m telescope with an angular resolution of about 4.3 arcmin and an rms noise of~0.05 K (F\"{u}rst \etal1990), was also analyzed.\\ 
\indent Images in the mid and far infrared obtained with the MSX and IRAS satellites were used to investigate the dust emission distribution.  The MSX data consist of images at \hbox{8.3$\,\umu$m}
(Band A),  \hbox{12.1$\,\umu$m} (Band C), \hbox{14.7$\,\umu$m} (Band D), and \hbox{21.3$\,\umu$m} (Band E), obtained with an angular resolution of 18\farcs3. The IRAS (HIRES)\footnote{IPAC is funded by NASA as part of the {\it IRAS} extended mission under contract to Jet Propulsion Laboratory (JPL) and California Institute of Technology (Caltech).} data include images at 12, 25, 60, and 100$\,\umu$m with angular resolutions in the range 0\farcm5 to 2\farcm0. The IRAS images are included in the DRAO dataset.

$^{12}$CO \hbox{(J =1 $\rightarrow$ 0)} data from Dame \etal(2001) were used to analyze the molecular gas distribution in the region. These data have an angular resolution of 7.5 arcmin, a velocity resolution of 1.3 \kms, and an rms noise of 0.05 K.

\begin{table}
\caption{DRAO data: Observational Parameters.}
\vspace{1.5mm}
\begin{tabular}[h!]{lc}
\hline\hline
Parameter&Value\\
\hline\hline
\multicolumn{2}{c}{\hi}\\
Synthesized beam&1\farcm22~$\times$~0\farcm97\\
rms noise (single channel) [K]&1.0\\
Bandwith [MHz]& 1.0\\
Channel separation [\kms]& 0.82\\
Velocity resolution  [\kms]& 1.3\\
Velocity range [\kms]& --165, +45\\
\\
\multicolumn{2}{c}{1420~MHz}\\
Synthesized beam&1\farcm03~$\times$~0\farcm96\\
rms noise [K]&0.06\\
Bandwith [MHz]& 30.0\\
\\
\multicolumn{2}{c}{408~MHz}\\
Synthesized beam&3\farcm5~$\times$~2\farcm8\\
rms noise [K]&0.4\\
Bandwith [MHz]& 4.0\\
\hline\hline
\end{tabular}
\end{table}

\section{The gas and dust distributions}

\subsection{The ionized gas}

Figure 2 shows the radio continuum emission distributions at~408~MHz and 1420~MHz (upper panels), and overlays of the radio continuum emissions and the VTSS image (lower panels). The images at both frequencies show emission from LBN\,148.11--0.45, and from the shell-like features at \hbox{$\sim${\it (l,b)} = (148\degr 50\arcmin, --0\degr 40\arcmin)} and \hbox{$\sim${\it (l,b)} = (149\degr 15\arcmin, --0\degr)}.

\begin{figure*}
\centering
\label{continuo}
\includegraphics[angle=0,width=0.8\textwidth]{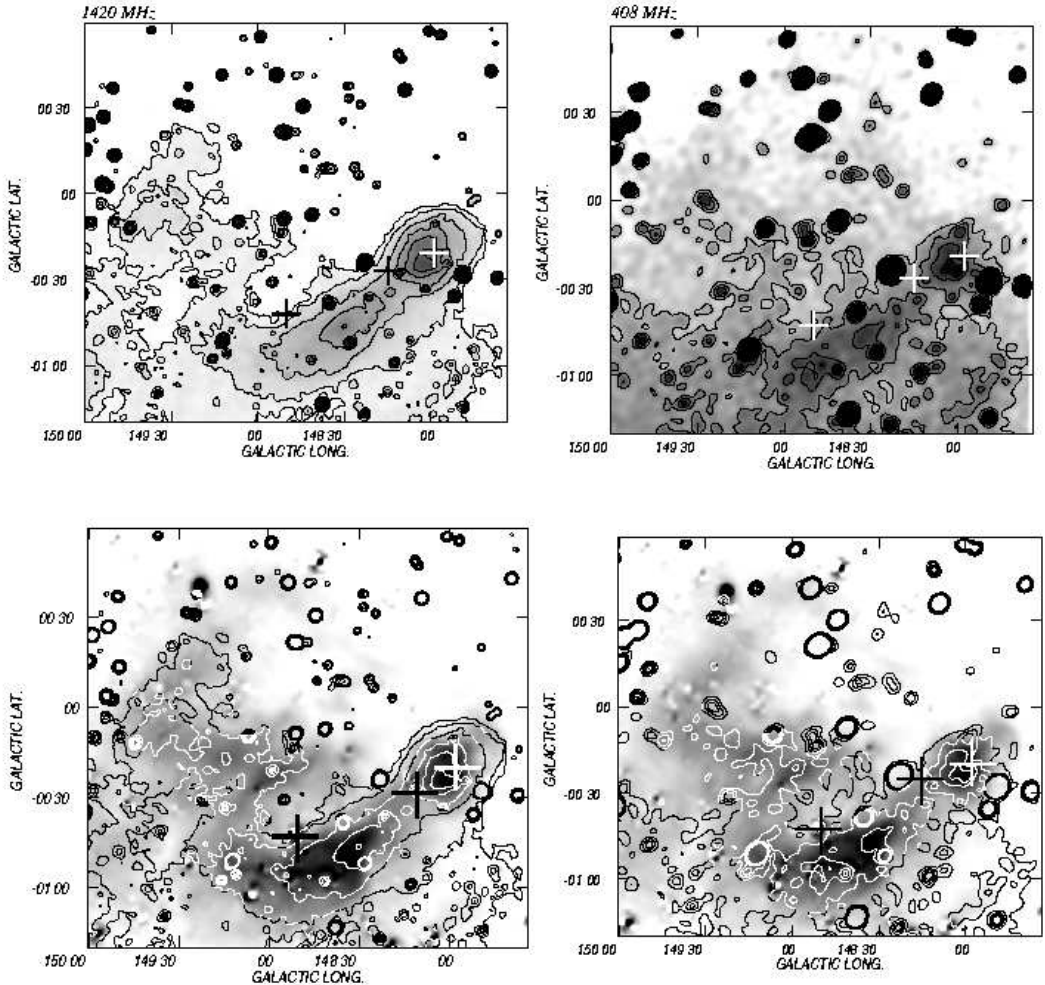}
\vspace{-1cm}
\caption{\small{Radio continuum images at 1420~MHz and 408~MHz. {\it Top left panel}: Image at 1420~MHz. Contour levels are from 5.2 to 6.1 K in steps of 0.2 K. {\it Top right panel}: Image at 408~MHz. Contour levels are from 53 to 61 K in steps of 2 K. {\it Bottom left panel}: Overlay of the 1420~MHz image (contours) and the VTSS image. {\it Bottom right panel}: Overlay of the 408~MHz image (contours) and the VTSS image.}}
\end{figure*}

The 1420~MHz radio continuum emission associated with LBN\,148.11--0.45 is strong. It consists of an almost circular, centrally picked region, of  \hbox{$\sim$ 30\arcmin $\times$ 24\arcmin} in size. The radio peak at \hbox{$\sim${\it (l,b)} = (147\degr 59\farcm4, --0\degr 25\arcmin)} does not coincide with the maximum in optical emission at \hbox{$\sim${\it (l,b)} = (147\degr 58\arcmin, --0\degr 19\arcmin)}.

As regards SH\,148.83--0.67 (related to HD\,24431), the radio emission at 1420~MHz is smooth and very well defined, with the brighter optical region coincident with the highest radio continuum emission at \hbox{$\sim${\it (l,b)} = (148\degr 30\arcmin, --0\degr 46\arcmin)}. The borders of the optical arc are brighter than the inner region, with the radio image displaying a rather uniform region instead. This radio emission extends towards the galatic plane, appearing as a diffuse and faint almost complete ring of 13\arcmin~in width, and 60\arcmin~in diameter, which correlates with the diffuse optical emission in the shell. Due to the lower angular resolution of the image at~408~MHz, the ring is not as well defined as at 1420~MHz.  HD\,24431 is seen projected onto the inner border of this ring, near one of the bright optical emission regions. 

\indent Diffuse radio continuum emission also correlates  with SH\,149.25--0.00. It is not clear if the two shell-like features, namely \hbox{SH\,149.25--0.0} and \hbox{SH\,148.83--0.67}, are linked.

\begin{table*}
\centering
\caption{\small{Small diameter sources projected onto the region.}}
\small
\vspace{3mm}
\begin{tabular}{cccccc}
\hline\hline
\multicolumn{2}{c}{Position }& S$_{1420}$ &S$_{408}$  & $\alpha$&Source\\
{\it l} [$\degr$] & {\it b} [$\degr$] & [mJy]  & [mJy]  & & designation\\
 \hline
\multicolumn{6}{c}{sources seen onto LBN\,148.11--0.45}\\
 \hline
147\degr 49\arcmin&--0\degr 28\arcmin&503 $\pm$ 0.93&1038 $\pm$ 18 &--0.58 $\pm$ 0.03&\footnotesize{NVSS~J035125+532828}
  \\
147\degr 52\arcmin&--0\degr 35\arcmin&80 $\pm$ 0.92&167 $\pm$ 17 &--0.59 $\pm$  0.17&\footnotesize{NVSS~J035107+532041}
  \\
147\degr 54\arcmin&--0\degr 25\arcmin&13.3 $\pm$ 0.91&14 $\pm$ 10 &--0.03 $\pm$ 1.25&\footnotesize{NVSS~J035205+532753}
  \\
148\degr 02\arcmin&--0\degr 33\arcmin&19.9 $\pm$ 0.89&39 $\pm$ 21 &--0.54 $\pm$ 0.92&\footnotesize{NVSS~J035210+531621}
  \\
148\degr 11\arcmin&--0\degr 28\arcmin&21.2 $\pm$ 0.84&60 $\pm$ 32 &--0.8 $\pm$ 0.2&\footnotesize{NVSS~J035320+531453}
  \\
147\degr 59\arcmin&--0\degr 11\arcmin&13 $\pm$ 0.9&45 $\pm$ 27 &--1.0 $\pm$ 0.2&\footnotesize{NVSS~J035327+533601}
  \\
  \hline
\multicolumn{6}{c}{sources seen onto SH\,148.83--0.67}\\
\hline
148\degr 23\arcmin&--0\degr 24\arcmin&405.4 $\pm$ 1.5&1288 $\pm$ 19 &--0.92 $\pm$ 0.02&\footnotesize{NRRF~J035430.6+530759}
   \\
148\degr 35\arcmin&--0\degr 39\arcmin&59.2 $\pm$ 1.6&177 $\pm$ 20 &--0.88 $\pm$ 0.16&\footnotesize{6C~B035049.0+524145}
  \\
148\degr 37\arcmin&--0\degr 37\arcmin&17.8 $\pm$ 1.5&149 $\pm$ 62 &--1.7 $\pm$ 0.6&\footnotesize{NVSS~J035337+530130}
  \\
148\degr 28\arcmin&--0\degr 52\arcmin&70.9 $\pm$ 1.6&81 $\pm$ 19 &--0.12 $\pm$ 0.38&\footnotesize{87GB~034917.6+523604}\\
148\degr 22\arcmin&--0\degr 47\arcmin&9.2 $\pm$ 1.8&17 $\pm$ 8 &--0.48 $\pm$ 0.57&\footnotesize{NVSS~J035256+525159}
  \\
148\degr 16\arcmin&--0\degr 58\arcmin&24.9 $\pm$ 1.6&14 $\pm$ 5 & +0.47 $\pm$ 0.51&\footnotesize{NVSS~J035651+521521}
   \\
149\degr 12\arcmin&--0\degr 51\arcmin&20.4 $\pm$ 1.5&419 $\pm$ 19 &--0.57 $\pm$ 0.074&\footnotesize{87GB~035303.4+520859}  \\
148\degr 51\arcmin&--0\degr 09\arcmin&123.4 $\pm$ 1.5& \small{not resolved} &&\footnotesize{87GB~035414.7+525443}\\
 148\degr 52\arcmin&--0\degr 14\arcmin&5.06 $\pm$ 0.32&18 $\pm$ 5&--1.01 $\pm$ 0.34&\footnotesize{NVSS~J035748+525933}\\
 148\degr 59\arcmin&--0\degr 21\arcmin&15.7 $\pm$ 1.5&35 $\pm$ 18 &--0.64 $\pm$ 0.2&\footnotesize{NVSS~J035751+524924}
  \\
148\degr 41\arcmin&--0\degr 58\arcmin&1.94 $\pm$ 0.25&33 $\pm$ 16 &--2.27 $\pm$ 0.68&\footnotesize{NVSS~J035347+523208}
  \\
148\degr 13\arcmin&--0\degr 59\arcmin&30.3 $\pm$ 1.9&30 $\pm$ 7 &+0.01 $\pm$ 0.36&\footnotesize{NVSS~J035115+525046}
  \\
\hline\hline
\end{tabular}\\
Notes:\\
Col. 6: References found in NED catalogue:
6C: Hales, Baldwin \& Warner (1988),
NVSS: Condon et al.~(1998),
87GB: Gregory \& Condon~(1991),
NRRF: Newberg et al.~(1999)\\
\end{table*}

\normalsize
As a first step to derive the physical parameters of the extended radio structures, we have analyzed the small diameter
sources projected onto and near the optical emission regions. Their galactic coordinates and derived flux densities measured at 1420~MHz and 408~MHz are listed in Table\,3. Their spectral indices {\it $\alpha$} (S$_{\nu}$~$\varpropto$~$\nu^{\alpha}$) derived using fluxes at these frequencies are shown in the fifth column. The sixth column indicates a previous reference to the sources. The derived spectral indices indicate that all but two sources are non-thermal in nature. These results are reinforced by the fact that all sources are identified as extragalactic sources in the {\it NASA Extragalactic Database} (NED). Two sources are particularly interesting. One of them, $\hbox{NRRF  J035430.6+530759}$, is a galaxy. The other one, \hbox{NVSS J035347+523208}, has a steep spectral index \hbox{(-2.27)}, which could correspond to a pulsar. However, this source is not listed in pulsars' catalogues (see for example Taylor, Manchester \& Lyne~(1993).

Flux densities at 2700~MHz, 1420~MHz, and 408~MHz for LBN\,148.11--0.45 and SH\,148.83--0.67, along with angular and linear sizes, are listed in Table\,4. Flux densities corresponding to the small diameter non-thermal sources were substracted from the measured fluxes at 1420~MHz, 408~MHz, and 2700~MHz for the two nebulae. The expected flux density of the non-thermal sources at 2700~MHz was estimated from the spectral index. The spectral indices derived using the resulting flux densities at 408, 1420, and 2700~MHz are listed in Table\,4, indicating that the emission originates in thermal bremsstrahlung. We note that the emission at 2700~MHz is very low and may lead to a large uncertainty in the estimate of the spectral index. By including the flux densities at 408 and 1420~MHz only, the spectral index results $\alpha$ = +0.14~$\pm$~0.02 for SH\,148.83--0.67 and +0.4~$\pm$~0.04 for LBN\,148.11--0.45, also compatible with thermal nature. Thus, the derived spectral indices show that the sources are thermal in nature.

\begin{table}
\begin{center}
\small
\caption{Physical parameters of \hii~regions.}
\vspace{3mm}
\begin{tabular}{lc}
\hline\hline
\multicolumn{2}{c}{SH\,148.83--0.67}\\
\hline\hline
Related star&HD\,24431\\
Size (\arcmin)&60~$\times$~54\\
Adopted distance (kpc)& 1.0~$\pm$~0.2\\
Size (pc) & 17.5~$\times$~15.7 \\
Radius (pc)& 8.3\\
{\it S}$_{1420}$ (Jy) &7.2~$\pm$~0.2\\
S$_{408}$ (Jy)  & 6~$\pm$~0.3\\
{\it S}$_{2700}$ (Jy)  &3.7~$\pm$~1.8\\
$\alpha$&--0.20~$\pm$~ 0.12\\
T$_{e}$ (K)&8000~$\pm$~2000\\
EM ($pc~cm^{-6}$)&520~$\pm$ 120\\
\multicolumn{2}{c}{{\it f}~=~1}\\
{\it n$_{e}$} (cm$^{-3}$)&6~$\pm$~1\\
M$_{i}$ (M$_{\sun}$)&420~$\pm$ 170\\
\multicolumn{2}{c}{{\it f}~=~0.1--0.3}\\
n$_{e}$ (cm$^{-3}$)& 15~$\pm$~4\\
M$_{i}$ (\msun)& 180~$\pm$~ 90\\
\hline
\hline
\multicolumn{2}{c}{LBN\,148.11--0.45}\\
\hline
Related stars&HD\,24094, ALS\,7793\\
Size (\arcmin)&30~$\times$~24\\
Adopted distance (kpc)& 1.0~$\pm$~0.2\\
Size (pc)&8.7~$\times$~7.0\\
Radius (pc)& 3.9\\
{\it S}$_{1420}$ (Jy) &3.2~$\pm$~0.1\\
{\it S}$_{408}$ (Jy)  & 1.90 ~$\pm$~0.07\\
{\it S}$_{2700}$ (Jy)  & 2.3~$\pm$~1\\
$\alpha$&+0.13~$\pm$~0.1\\
T$_{e}$ (K)&10000\\
EM ($pc~cm^{-6}$)&1000~$\pm$ 160\\
\multicolumn{2}{c}{{\it f}~=~1}\\
{\it n$_{e}$} (cm $^{-3}$)&12 ~$\pm$~1\\
M$_{i}$ (M$_{\sun}$)&70~$\pm$ 25 \\
\hline\hline
\end{tabular}
\end{center}
\end{table}

\normalsize
The physical parameters of the H{\sc ii} regions, namely, emission measure {\it EM}, electron density {\it n$_{e}$}, and ionized mass {\it M$_{i}$}, were obtained using the expressions by Mezger \& Henderson (1967) and correspond to a filling factor  f\,=\,1.
For the case of SH\,148.83--0.67, a more realistic volume filling factor was estimated considering a shell-structure with inner radius R$_{in}$ and outer radius R$_{ou}$. Both radii were measured using the radio continuum map at 1420~MHz, taking into account the contour level of 5.15 K ($\equiv$ 85 $\sigma$) (see Fig.\,2). The values 28\arcmin~and 13\arcmin~were obtained for R$_{ou}$ and R$_{in}$, respectively. For the adopted distance, {\it d} = 1 kpc, R$_{ou}$ = 8.2 pc and R$_{in}$= 3.8 pc. Assuming that \hbox{10--30 per cent} of the shell surface is occupied by plasma, the derived volume filling factor is in the range 0.1--0.3.

The assumed electron temperature is \hbox{10\,$^{4}$ K} for the case of LBN\,148.11--0.45, and values in the range \hbox{(0.6-1.0) $\times$ 10\,$^{4}$  K} were used for SH\,148.83--0.67. The derived ionized masses were multiplied by 1.27 to take into account singly ionized Helium (Goss \& Lozinskaya~1995). The results are listed in Table \,4. Uncertainties in masses and electron densities come mainly from the distance uncertainty. For the case of LBN\,148.11--00.45, the central radio peak suggests f = 1.

\subsection{Dust emission distribution}

Figure 3 illustrates the infrared emission distribution in the area under study. The top left panel displays the emission at \hbox{60$\,\umu$m} in contours and grayscale, while the top right panel shows an overlay of the infrared emission at \hbox{60$\,\umu$m} ({\it grayscale}) and the radio continuum emission at 1420 MHz ({\it contours}).

\begin{figure*}
\centering
\includegraphics[angle=0,width=0.9\textwidth]{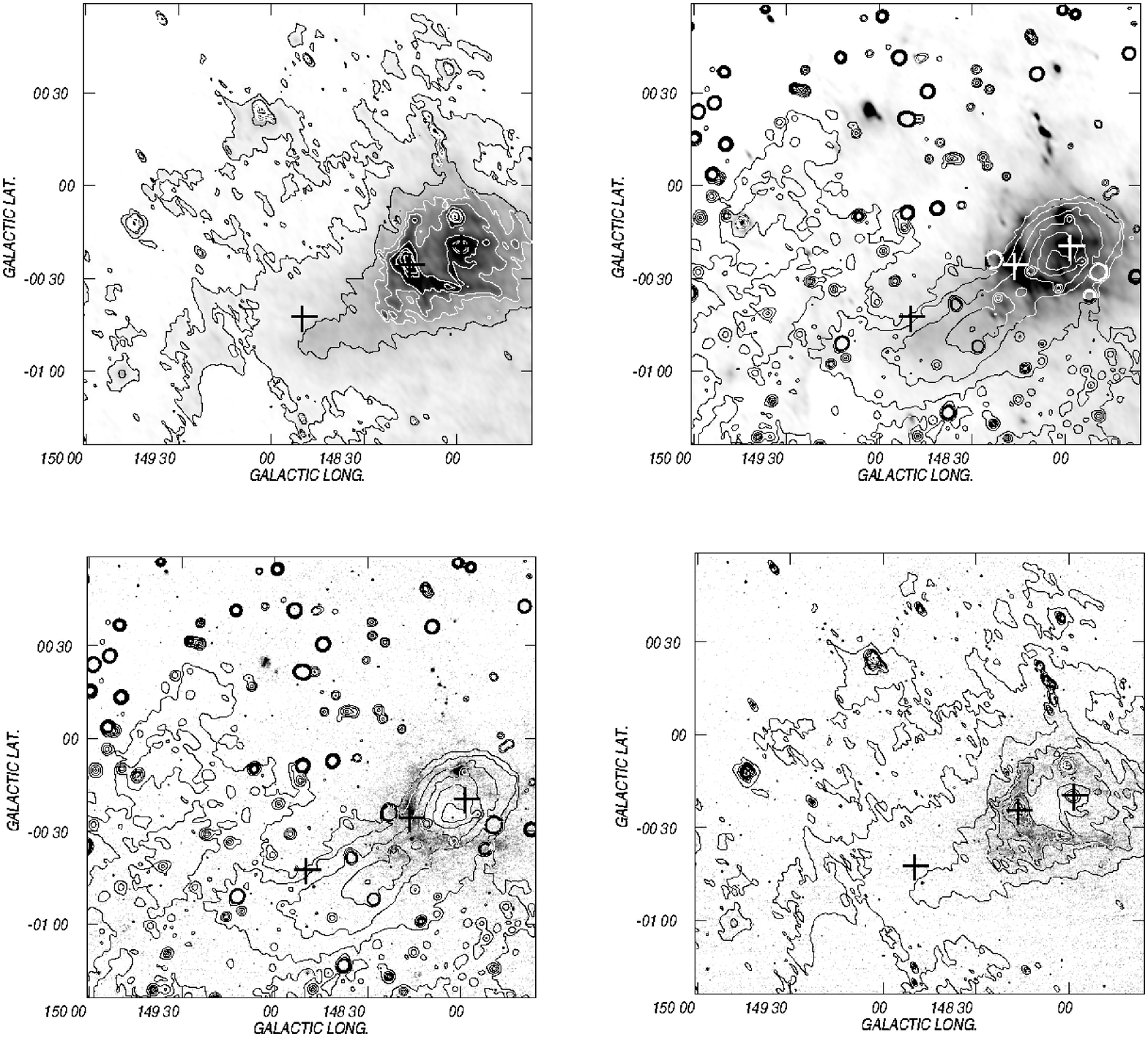}
\caption{\small{\it Upper left panel}: Image at \hbox{60$\,\umu$m} in grayscale and contour lines. Contour lines are 10, 15, 20, 25, 35, 50, 60, 100, and 140 MJY ster$^{-1}$. Grayscale is from 9.00 to 40.00 MJY ster$^{-1}$. {\it Upper right panel}: Overlay of the \hbox{60$\,\umu$m} image (grayscale) and 1420~MHz emission (contours). Contour levels are from 5.2 to 6.1 K in steps of 0.1 K. {\it Bottom left panel}: 1420~MHz emission (contours) superpossed onto the \hbox{8.3$\,\umu$m} image (grayscale). Grayscale corresponds to 600 $\times$ 10$^{-9}$ to 1400 $\times$ 10$^{-9}$ \hbox{W m$^{-2}$ ster$^{-1}$}. {\it Bottom right panel}: Overlay of the \hbox{8.3$\,\umu$m} image (grayscale) and \hbox{60$\,\umu$m} emission (contours). The crosses mark the positions of HD\,24431, ALS\,7793, and HD\,24094.}
\label{1420-60-8}
\end{figure*}

\indent The emission distribution in the far infrared differs from the radio continuum emission. Strong emission comes from \hbox{LBN\,148.11--0.45}. Arc-shaped extended emission at  \hbox{60$\,\umu$m} strikingly surrounds the border of the \hii~region towards larger galactic longitudes and separates this region from  \hbox{SH\,148.83--0.67}. 

Two small emission regions are easily distinguished in \hbox{LBN\,148.11--0.45}. One of them is a relatively extended feature coincident with the stellar position. This \hbox{60$\,\umu$m} maximum is shifted  5 arcmin from the maximum at 1420 MHz. The other infrared region is detected towards \hbox{{\it (l,b)} = (148\degr 0\arcmin, --0\degr 10\arcmin)}. Its position is coincident with the small diameter source NVSS~J035327+533601, detected in the radio continuum and suspected to be an extragalactic source.

\indent Faint diffuse infrared emission also overlaps part of \hbox{SH\,148.83--0.67}. This weak feature coincides with the arc-like structure seen at 1420 MHz and at optical wavelengths (see Fig. \ref{optico}). No far infrared emission is identified near \hbox{{\it (l,b)} = (149\degr 00\arcmin, --0\degr 20\arcmin)}, where faint optical emission belonging to \hbox{SH\,148.83--0.67} was detected.

\indent The image at \hbox{100$\,\umu$m} is not shown here since the emission distribution is similar to that at \hbox{60$\,\umu$m}.

The bottom right panel of Fig. 3 shows an overlay of the infrared emission at \hbox{8.3$\,\umu$m} (Band A of MSX satellite) ({\it grayscale}) and the infrared emission at \hbox{60$\,\umu$m} ({\it contours}). The bottom left panel shows an overlay of the infrared emission at \hbox{8.3$\,\umu$m} ({\it grayscale}) and the radio continuum emission at 1420 MHz ({\it contours}). The image at \hbox{8.3$\,\umu$m} reveals an arc-shaped feature, 5\arcmin~in width, which delineates the border of the \hii~region towards larger galactic longitudes and lower galactic latitudes, and coincides with a similar feature seen at \hbox{60$\,\umu$m}. No emission in the MSX band A is detected close to HD\,24431. Very likely, the image at \hbox{8.3$\,\umu$m} shows radiation from polycyclic aromatic hydrocarbons ({\it PAHs}), which can survive in the neutral gas around \hii~regions (Hollenbach \& Tielens~1997), but not inside the ionized regions. Thus, this emission traces a PDR lying in the outskirts of \hbox{LBN\,148.11--0.45}.

NVSS~J035327+533601 also has a bright counterpart at \hbox{8.3$\,\umu$m}. No significant emission at \hbox{8.3$\,\umu$m} is detected in the rest of the area. We note that band E is free of emission in the whole area.

The IRAS (HIRES) image at \hbox{12$\,\umu$m} (not shown here) presents the same spatial  distribution as the \hbox{8.3$\,\umu$m} image, since the filter responses of A and \hbox{12$\,\umu$m} bands are similar (Egan~1996).  These bands are detecting dust with the same physical conditions.

\begin{table*}
\centering
\caption{\small{Parameters of the interstellar dust associated with LBN\,148.11--0.45 and SH\,148.83--0.67.}}
\small
\vspace{3mm}
\begin{tabular}{lcccccc}
\hline\hline
\hii~region&S$_{\hbox{60$\,\umu$m} }$&S$_{\hbox{100$\,\umu$m} }$&T$_{d}$ (K)&\multicolumn{2}{c}{M$_{d}$ (\msun) }\\
             ~      & (Jy)& (Jy)& (n =1.0--2.0)& n = 1.0 & n= 2.0\\
	     \hline
LBN\,148.11--0.45 ({\bf a}) & 680   &  1410  &  32 $\pm$ 3       & 0.2 $\pm$ 0.1& 0.8 $\pm$ 0.3 \\
LBN\,148.11--0.45 ({\bf b}) &  910 &   2470 &  29 $\pm$ 3      &0.6 $\pm$ 0.3& 2.0 $\pm$ 0.9\\
LBN\,148.11--0.45 ({\bf c})&   260  &  740  &  29  $\pm$ 3    & 0.2 $\pm$ 0.1&0.6 $\pm$ 0.3\\
LBN\,148.11--0.45 ({\bf d})& 2900 $\pm$ 100&   8100 $\pm$ 300  & 29 $\pm$ 3 &0.4  $\pm$   0.2 & 1.5 $\pm$ 0.6\\
SH\,148.83--0.67 & 620 $\pm$ 180   &1750 $\pm$ 450 & 29  $\pm$ 3    &2.0 $\pm$ 0.8& 7.0 $\pm$ 3.0\\
\hline
\end{tabular}
\end{table*}

Table\,5 lists flux densities at \hbox{60$\,\umu$m} and \hbox{100$\,\umu$m}, dust temperatures, and dust masses associated with \hbox{LBN\,148.11--0.45} and \hbox{SH\,148.83--0.67}. The parameter $n$ is related to the dust absorption efficiency ($\kappa_{\nu}$~$\propto$~$\nu^{n}$). The dust temperature was estimated for $n$ in the range 1 to 2. The infrared emission related to \hbox{LBN\,148.11--0.45} was split in three regions to obtain the physical parameters of the associated dust. The first region involves the infrared emission which correlates with the central part of the \hii~region ({\bf a}), the second one is the arc-shaped structure ({\bf b}), and the third one is the extended emission coincident with NVSS~J035327+533601 ({\bf c}). Region ({\bf d}) corresponds to the \hii~region as a whole. Uncertainties in listed dust masses originate in the adopted distance error and in different infrared background {\bf emissions}. No significant difference in dust color temperature is apparent among the different sections of \hbox{LBN\,148.11--0.45}. Derived dust temperatures for \hbox{SH\,148.83--0.67} and \hbox{LBN\,148.11--0.45} are typical for \hii\ regions. 

\subsection{The \hi~gas emission distribution}

To find possible \hi~structures associated with the ionized regions, the \hi~data cube was carefully examined within the velocity range from --165 to +45 \kms~(all velocities in this paper are with respect to the LSR).

\begin{figure}
\centering
\includegraphics[angle=0,width=8cm]{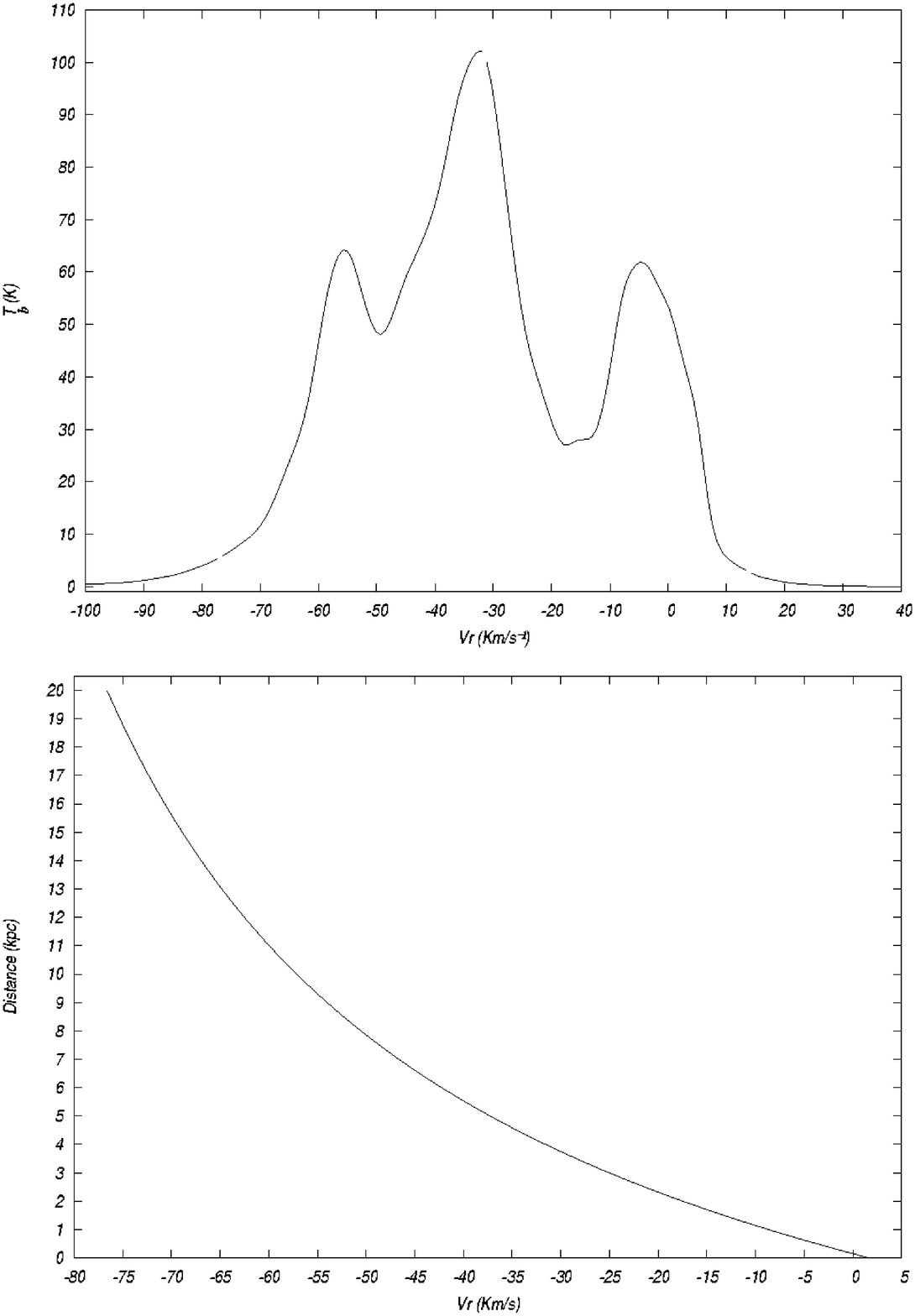}
\caption{\small{{\it Top panel}: Average \hi~profile within a region of 2\degr $\times$ 2\degr~in size, centered at \hbox{{\it (l,b)}=(148\degr29\arcmin, --0\degr19\arcmin)}. {\it Bottom panel}: Kinematical distance vs. Radial velocity plot derived from the analytical fit to the rotation curve by Brand \& Blitz~(1993).}}
\label{perfil-HI}
\end{figure}

The average \hi~spectrum corresponding to a region of 2\degr$\times$2\degr~in size centered at \hbox{{\it (l,b)} = (148\degr29\arcmin, --0\degr19\arcmin)} is displayed in Fig. 4 to facilitate visualization of the large scale characteristics of the \hi~in the area. The \hi~profile shows the presence of significant emission covering the velocity range from --80 to +10 \kms. Three main \hi~components can be distinguished. According to circular galactic rotation models (Brand \& Blitz~1993), components centered at +4 \kms, --32 \kms~and \hbox{--56 \kms}~should be located at kinematical distances \hbox{$\la$ 1 kpc}, 4 $\pm$ 1 kpc and 9 $\pm$ 2 kpc, respectively. These features are probably related to emission from the local and the Perseus spiral arms (Georgelin \& Georgelin~1976).

The \hi~profile presents a minimum at \hbox{$\approx$ --15 \kms}. The circular galactic rotation model predicts a kinematical distance of $\approx$ 1.6 $\pm$ 0.8 kpc for gas at this velocity, compatible with distance estimates for HD~24431 (see Table\,1). In order to look for cavities and holes that may be associated with HD~24431 and HD~24094 we have meticulously inspected  the velocity interval \hbox{[+6,--50]} \kms.

\begin{figure*}
\includegraphics[angle=0,width=\textwidth]{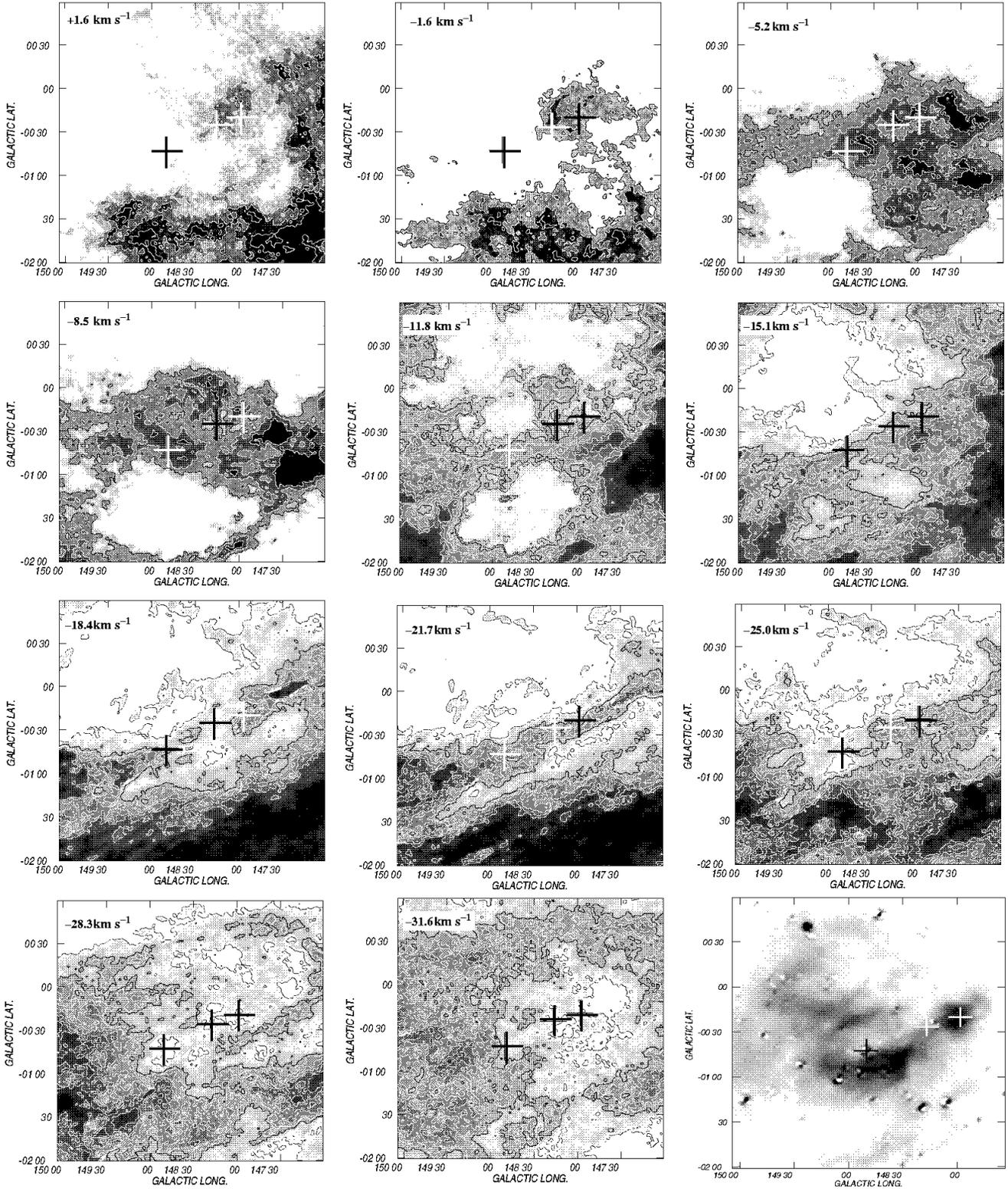}
\caption{\small{\hi~emission distribution in the velocity range from +1.6 to --31.6 \kms. The crosses mark the positions of HD~24431, ALS\,7793, and HD\,24094. The central velocity of each map is indicated in its upper left corner. Contour levels for maps within the velocity interval [+1.6, --11.8] \kms~are from 4  to 16  K in steps of 4 K. Contour levels for maps within the velocity interval [--15.1, --31.6] \kms~are from --8  to 16  K in steps of 4 K. The grayscale for maps having velocities of +1.6 \kms~and --1.6 \kms~are from 0 K to 10 K. The grayscale for  the rest of the maps is from --10 K to 30 K. The optical image of Sh2-205 is shown in the bottom right corner. Darker regions mean  brighter areas.}}
\end{figure*}

Figure 5 shows the \hi~emission within the velocity interval +1.6 to --31.6 \kms~in steps of 3.3 \kms. Each image is the average of four consecutive \hi~maps. In order to facilitate the display of the images, a constant background, equal to the mean value of each image within a box of 2\degr $\times$ 2\degr~in size centered at {\it (l,b)} = (148\degr29\arcmin, --0\degr19\arcmin) was substracted from every map. The \hi~gas distribution is quite complex and, as a consequence, the identification of the neutral gas associated with the ionized regions is hard to establish. We will discuss each case separately, along with an \hi~shell centered at \hbox{{\it (l,b)}=(149\degr0\arcmin, --1\degr30\arcmin)}.

\subsubsection{\hi~gas related to LBN\,148.11--0.45}

An inspection of Fig. 5 shows that \hi~gas  related to LBN\,148.11--0.45 is detected within the velocity range from +1.6 to --15.1 \kms. At +1.6 \kms, a region of enhanced \hi~emission is seen partially coincident with the \hii~region. This feature consists of a faint elongated \hi~cloud centered at {\it (l,b)} = (148\degr 10\arcmin, --0\degr 15\arcmin) that extends 30 arcmin~at a position angle of $\approx$ 45\degr. Within the velocity interval from --1.6 to --5.2 \kms, \hi~gas is clearly  seen surrounding the \hii~region near \hbox{{\it (l,b)} $\thickapprox$ (148\degr 15\arcmin, --0\degr 15\arcmin)}. In the velocity range from \hbox{--5.2 to --11.8} \kms, a patchy envelope encircles the \hii~region almost completely. At \hbox{$\approx$ --8.5 \kms},  the brightest section of the \hi~structure surrounds the \hii~region towards {\it b} $\sim$ --0\degr 40\arcmin. The envelope is hardly identifiable at \hbox{v $\la$ --15.0 \kms}. At \hbox{v $\thickapprox$ --22.0 \kms}, \hbox{HD\,24094} and \hbox{ALS\,7793} are seen projected onto a bright \hi~filament that runs along a position angle of $\approx$ 30\degr~across the area.

Figure 6 shows the \hi~emission in the velocity range from --1.5 to --11.8 \kms~(left panel), and an overlay of the 1420~MHz continuum emission and the \hi~image (right panel). \hi~gas most probably related to \hbox{LBN\,148.11--0.45} is clearly seen in the left panel, where the \hi~envelope is identified. We believe that the \hi~gas that encompasses \hbox{LBN\,148.11--0.45} represents an atomic gas shell associated with the ionized region. \hi~gas projected onto the central part of the \hii~region, which is detected within the velocity interval +1.5 to --5.2 \kms, may represent part of the receding cap of the envelope.

\begin{figure*}
\label{hI-1-3}
\centering
\includegraphics[angle=0,width=0.7\textwidth]{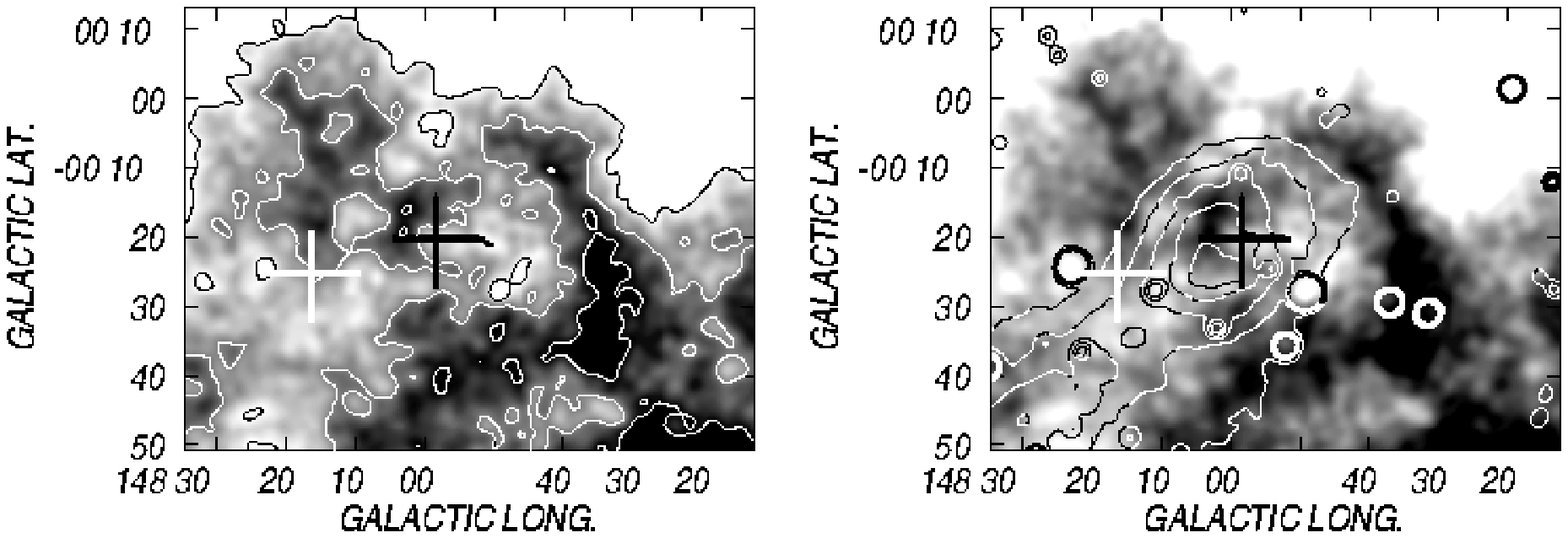}
\caption{\small {\it Upper panel}: \hi~emission distribution towards \hbox{LBN\,148.11--0.45} within the velocity range --1.5 to --11.4 \kms. The crosses indicate the positions of ALS\,7793 and 
HD\,24094. The grayscale is from 60 to 70 K. Contour levels are from 60 to 75 in steps of 5 K. {\it Lower panel}: Overlay of the \hi~emission shown in the left panel ({\it grayscale}) and the radio continuum image at 1420~MHz ({\it contour lines}). Contour levels are from 5.3 to 6.1 K in steps of 0.2 K. Darker regions mean brighter areas.}
\end{figure*}

\subsubsection{\hi~gas related to SH\,148.83--0.67 and \hbox{SH\,149.25--0.0}}\label{HI-gs}

Identification of neutral atomic gas related to these features is particularly difficult.

Figure\,5 shows a low emission region surrounded by an almost complete shell centered at {\it (l,b)} = (148\degr50\arcmin, --0\degr35\arcmin), which can be identified in the velocity range from --22.0 to --32.0 \kms.

Figure\,7 shows the \hi~emission averaged in the velocity range from --25.0 to --28.0 \kms, where this feature is better defined. The cavity and the outer shell, 1\fdg8 $\times$ 1\fdg3 in size, are shown in the left panel. The borders of the cavity are defined by the contour line corresponding to 70 K at \hbox{{\it b} $\le$ --0\degr 30\arcmin}~and by the contour line of 60 K at \hbox{{\it b} $\ge$ --0\degr 30\arcmin}. The right panel, which displays a superposition of the \hi~and optical images, reveals that the optical arc-like feature at \hbox{{\it (l,b)} = (148\degr 45\arcmin, --0\degr 50\arcmin)} appears projected onto the cavity. The faint optical emission at \hbox{{\it (l,b)} = (149\degr, --0\degr 20\arcmin)} is seen within the cavity. The comparison between the \hi~and optical emission suggests that this \hi~shell is related to  \hbox{SH\,148.83--0.67}.

\begin{figure*}
\centering
\includegraphics[angle=0,width=0.8\textwidth]{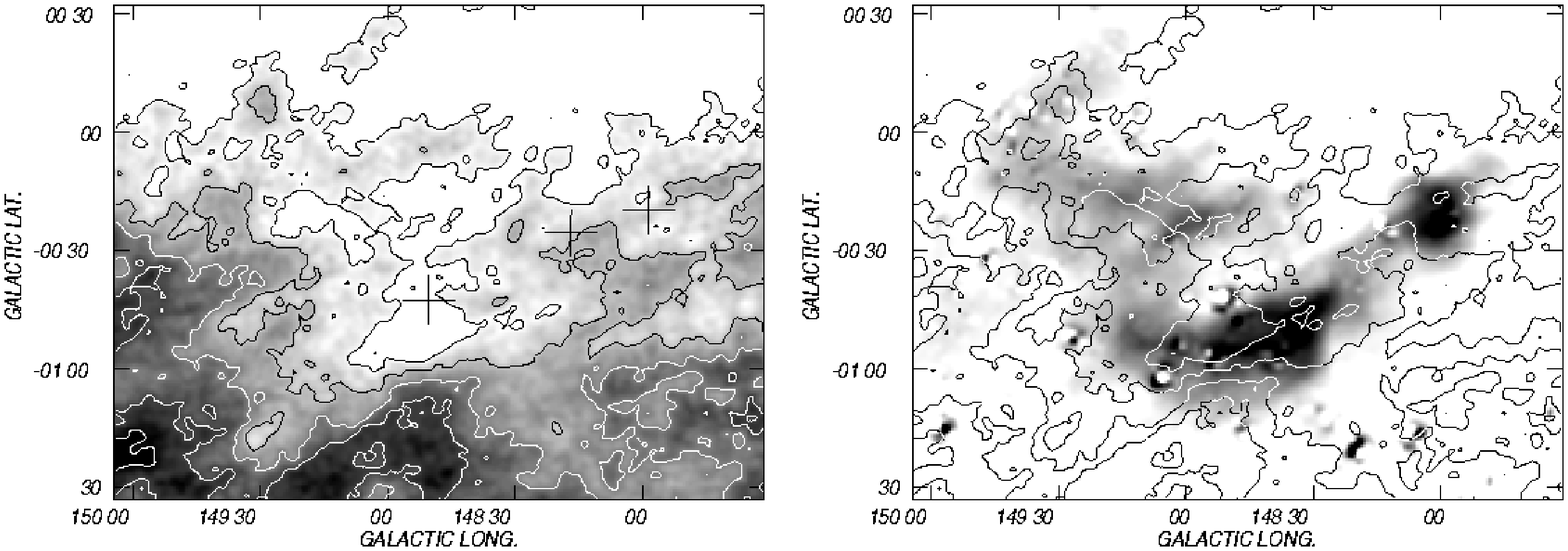}
\caption{\small {\it Left panel}: \hi~emission distribution for the structure probably related to SH\,148.83--0.67 within the velocity range \hbox{--25.0 to --28.0} \kms. The grayscale is from 60 to 100 K. The contour levels are 60 to 100 K in steps of 10 K. The cross symbols indicate the positions of HD\,24431, ALS\,773, and HD\,24094. {\it Right panel}: Overlay of the \hi~emission distribution ({\it contour lines}) and the VTSS image ({\it grayscale}). Darker regions mean brighter areas.}
\end{figure*}

A bright \hi~filament running along the complete interval of galactic longitudes at a position angle of 30\degr~can be easily distinguished at $\approx$ --22.0 \kms~(Fig. 5), with HD\,24431 projected onto the \hi~filament. Thin \hi~areas surround the bright knots seen in the optical image near {\it (l,b)} = (148\degr 40\arcmin, --0\degr45\arcmin) suggesting that \hi~gas at this velocity interacts with the nebula.\\
\indent For a systemic velocity of $\thickapprox$ --27 \kms, the circular galactic rotation model (see Fig.\,4) predicts a kinematical distance of  $\thickapprox$ 3.3 $\pm$ 0.8 kpc, assuming a velocity dispersion of 6 \kms. The derived distance is larger than spectrophotometric distance estimates for HD\,24431 (see Table\,1). Based on the spectrophotometric distance to this star, we adopt \hbox{{\it d} = 1.0 $\pm$ 0.2 kpc} as the distance to \hbox{SH\,148.83--0.67} and the surrounding \hi~shell.

\begin{figure}
\label{hI-3-3}
\centering
\includegraphics[angle=0,width=0.45\textwidth]{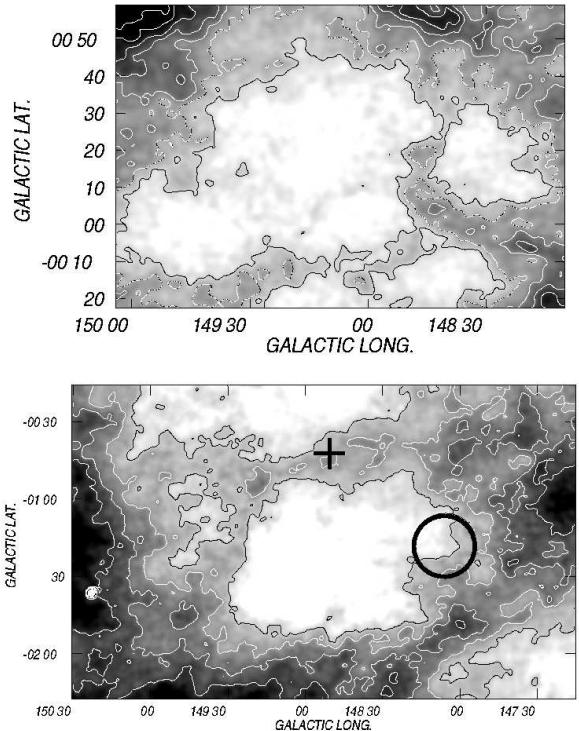}
 \caption{\small{\it Upper panel}: \hi~emission distribution for the neutral atomic gas structure probably related to SH\,149.25--0.0 within the velocity range --13.4 to --16.7 \kms. The grayscale is 15 to 45 K. The contour levels are 10, and from 20  to 50 K  in steps of 5 K. {\it Lower panel}: \hi~emission distribution within the velocity interval from --3.1 to --17.1 \kms~showing the \hi~structure at \hbox{{\it (l,b)} = (149\degr 0\arcmin, --1\degr 30\arcmin)}. The contour levels are from 50 to 65 K in steps of 5 K. The positions of HD\,24431 and NGC\,1444 are indicated by a cross symbol and a circle, respectively. Darker regions mean brighter areas.}
\end{figure}

Other \hi~structure in the area is centered at \hbox{{\it (l,b)} = (149\degr15\arcmin, +0\degr 25\arcmin)} at \hbox{v = -- 15.1 \kms}, of $\approx$ 60\arcmin~in size. The structure can be followed within a velocity range from \hbox{--12.0 to --18.4 \kms}. At $\approx$ --15 \kms, this feature presents its largest angular size, and its borders are better defined than at other velocities (Fig. 8, {\it upper panel}).  For gas at this velocity, the analytical fit by Brand \& Blitz~(1993) predicts a kinematical distance \hbox{{\it d} = 1.8 $\pm$ 0.8} kpc. The comparison with the radio continuum emission at 1420 MHz reveals that the weak radio emission at \hbox{{\it (l,b)} = (149\degr 15\arcmin, +0\degr20\arcmin)}, related to SH\,149.25--0.00, is seen projected onto the innermost part of the cavity, suggesting that the \hi~structure is linked to it. 

\subsubsection{An \hi~shell centered at \hbox{{\it (l,b)} = (149\degr0\arcmin, --1\degr30\arcmin)}}\label{HI-gs}

The analysis of the \hi~emission allows identification of an \hi~structure within the velocity range --5.2 to --15.1 \kms, centered at  \hbox{{\it (l,b)} = (149\degr 0\arcmin, --1\degr 30\arcmin)}, with a systemic velocity of $\thickapprox$ --12 \kms. The open cluster NGC\,1444, at a distance of 1.2 kpc (see Section 1), is placed near the border of the \hi~hole. The kinematical distance of the shell is compatible with the optical distance to the cluster. The structure is \hbox{$\approx$ 2\fdg2 $\times$ 1\fdg5}~in size and is shown in the lower panel of Fig.\,8.

\subsubsection{Physical parameters of the \hi~structures}

Table\,6 lists the main data of the neutral atomic structures identified in the region: the {\it (l,b)} centroid of the \hi~shells, the velocity interval spanned by
the structures, the systemic and expansion velocities, and the kinematical and adopted distances. The expansion velocity was estimated from the velocity range spanned by the \hi~feature as \hbox{(v$_{2}$ -- v$_{1}$) $/$ $2$ + $1.0$}. The size of the \hi\ structures was derived from the position of the maxima in the shells. The radius R$_{s}$, is defined as half the geometric mean of the axes of the \hi~structure. The neutral atomic mass corresponds to the mass excess in the shell. It was obtained within a circle of radius R$_{e}$, assuming that the gas is
optically thin, and includes a He abundance of \hbox{10 per cent}. The neutral atomic mass associated with the \hi~shell probably linked to \hbox{SH\,149.25--0.00} is difficult to establish.

\begin{table*}
\centering
\caption{\small{Parameters of \hi~and $^{12}$CO structures.}}
\small
\vspace{3mm}
\begin{tabular}{lcccc}
\hline
\hline
{\bf Related ionized regions}&{\bf LBN\,148.11--0.45}&{\bf SH\,148.83--0.67} &--&{\bf Sh\,149.25--0.0}\\
{\bf Neutral \hi~shells} &&&&\\
 ({\it l,b}) centroid  &147\degr 55\arcmin, --0\degr 20\arcmin &148\degr 50\arcmin, --0\degr 35\arcmin & 149\degr 0\arcmin, --1\degr 30\arcmin&  149\degr 15\arcmin, +0\degr 25\arcmin\\
Velocity interval v1,v2 (\kms)  &   +1.6, --18.0  &    --22,--32         &   --5,--15& --20, --12   \\
Systemic velocity (\kms)    &     --9~$\pm$~1  &     --27~$\pm$~1  &--12~$\pm$~1&--15~$\pm$~1     \\
V$_{exp}$ (\kms)   & 11 & 7& 7& 6 \\
Kinematical distance {\it d} (kpc) &           1.3 &       3.3            & 1.5  &  1.8\\
Adopted distance (kpc) & 1.0 $\pm$ 0.2& 1 $\pm$ 0.2&  1.0 $\pm$ 0.2 & 1.0 $\pm$ 0.2\\
Size of the \hi~structure (r$_{1}$ $\times$ r$_{2}$) (\arcmin)  &    51 $\times$ 49  &      105 $\times$ 80  &   129 $\times$ 90&    60 $\times$ 60  \\
Size of the \hi~structure (r$_{1}$ $\times$ r$_{2}$) (pc)        &  15 $\times$ 14    &   31 $\times$ 23  &  38 $\times$ 26 & 17 $\times$ 17  \\
Radius R$_{s}$  (pc)   &  7.2 &  13.4& 15.7&8.5\\
 R$_{e}$ (\arcmin) & 25   & 27    &    56 &--      \\
Neutral atomic mass (\msun)&   65 $\pm$ 25       &   300 $\pm$ 120  &1600$\pm$ 640 &--\\
\hline
{\bf Molecular gas} &&&&\\
Velocity interval v1,v2 (\kms) &{\bf --0.65},--11.1&--&--0.65,--11.1&--\\
W$_{{\rm CO}}$  (K \kms) &41 &-- &9 &--\\
Mean {\rm H$_{2}$} column density (10$^{21}$ mol$_{H_{2}}$ cm$^{-2}$)  &8 $\pm$ 0.8 & -- &2 $\pm$ 0.2&--\\
Mass (10$^{4}$ $\times$ \msun)&6.0 $\pm$ 3.0 &--&2.6  $\pm$ 1.3
&--\\
\hline
Ambient density (cm$^{-3}$)& $\sim$ 800    &$\sim$ 2              &$\sim$  70                         &\\
\hline\hline
\end{tabular}
\end{table*}

\subsection{The molecular gas emission distribution}

The analysis of the $^{12}$CO emission distribution in the region shows the presence of molecular gas associated with LBN\,148.11--0.45 within the velocity range from \hbox{--0.65 to --11.1 \kms}. Figure 9  (upper left panel) shows an overlay of the $^{12}$CO emission distribution within the velocity range \hbox{[--3.2, --4.5]} \kms~({\it contours}) and the infrared emission at 60 $\mu$m ({\it grayscale}). The molecular cloud at \hbox{{\it l} = (148\degr 25\arcmin)} delineates strikingly well the border of the PDR detected at \hbox{8.3$\,\umu$m}. Note that this CO cloud encompasses the brightest \hi~emission arc detected at \hbox{v = --1.6 \kms}~(Fig.\,5), suggesting that the thin \hi~filament can originate in the photodissociation of the molecular gas.

\begin{figure*}
\centering
\includegraphics[angle=0,width=0.7\textwidth]{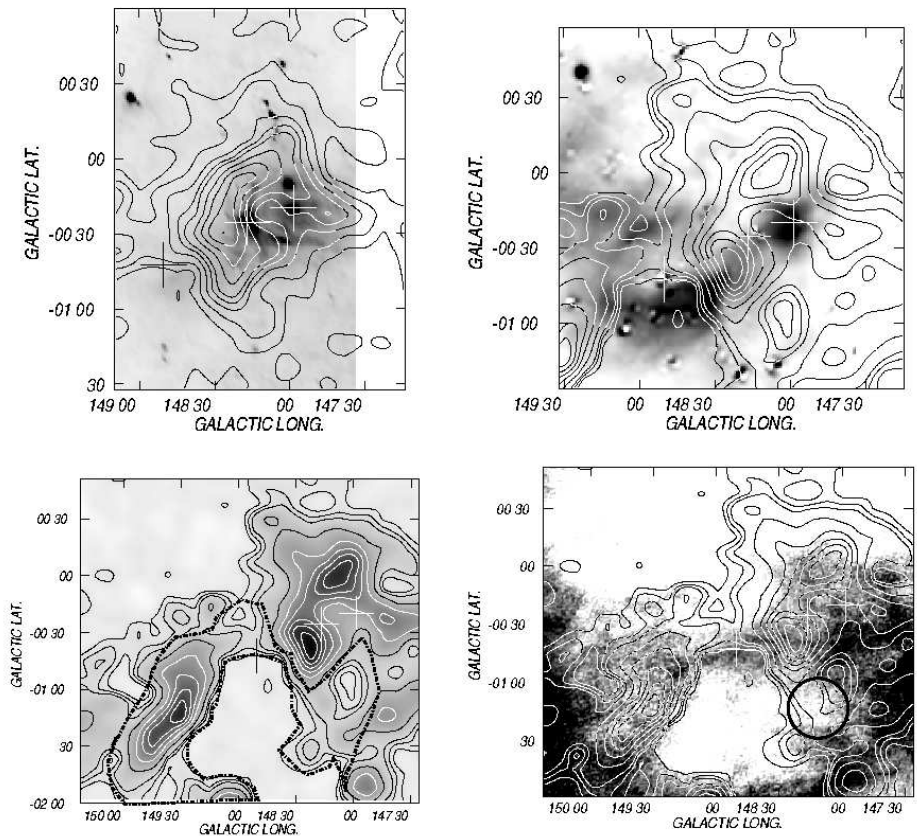}
\caption{\small{$^{12}$CO emission distribution. {\it Upper left panel}: Overlay of the integrated $^{12}$CO emission in the velocity range from --3.2 to --4.5 \kms~(contour lines) and the \hbox{60$\,\umu$m} image ({\it grayscale}). Contours levels are 0.15 ($\equiv$ 3$\sigma$), 0.5, 1, 1.5 K, from 2 to 6 K in steps of 1 K, and from 6 to 16 K in steps of 2 K. {\it Upper right panel}: $^{12}$CO emission integrated in the velocity inteval [--0.65, --11.1] \kms~superpossed onto the VTSS H$\alpha$ image ({\it grayscale}). {\it Bottom left panel}: $^{12}$CO emission emission integrated in the velocity inteval [--0.65, --11.1] \kms. The molecular mass was derived within the area marked out by the dash-dot line. {\it Bottom right panel}: Overlay the $^{12}$CO emission and the \hi~emission distributions within the velocity range from --3.1 \kms~to --17.1 \kms~({\it grayscale}). The contour levels indicated in the lower panels are the same that are displayed in the upper right panel. The positions of HD\,24431, ALS\,7793, and HD\,24094 are indicated by crosses.}}
\end{figure*}

CO emission seen towards the central part of the \hii~region \hbox{LBN\,148.11--0.45} is likely placed behind the ionized region. 

The upper right panel of Fig.\,9 shows the $^{12}$CO emission distribution in the velocity range [--0.65, --11] \kms~overimpossed onto the VTSS image. $^{12}$CO gas emission at \hbox{{\it (l,b)} = (148\degr25\arcmin, --0\degr40\arcmin)} coincides with the faint optical emission region which separates the optically bright regions \hbox{SH\,148.83--0.67} and \hbox{LBN\,148.11--0.45}.
The slight distorsion in the molecular clump at \hbox{{\it (l,b)} = (148\degr25\arcmin, --0\degr40\arcmin)} following the border of \hbox{SH\,148.83--0.67}, and the morphological correlation among the optical, radio continuum, and $^{12}$CO emissions indicates that the molecular gas is interacting with the ionized material.

The bottom panels of Fig.\,9 show the $^{12}$CO emission within the velocity interval --3.1 to --17.1 \kms~and a superposition of the CO and \hi~gas distributions in the region of the \hi~shell centered at \hbox{{\it (l,b)} = (149\degr0\arcmin, --1\degr30\arcmin)}. The morphological correspondence between the \hi~and CO emissions reveals that this \hi~shell has a clear molecular counterpart.

The open cluster NGC\,1444, marked in Fig. 9 by a circle,  is seen projected close to the border of the CO and \hi~emission rings. 

Towards the region under study, identification of molecular structures related to the ionized regions is easier than recognition of the associated \hi~structures, as can be expected bearing in mind confusion effects in the line of sight.

Table\,6 summarizes the physical parameters of the molecular clouds associated with the ionized and neutral regions. The $H_2$ column density ($N_{H2}$) was derived from the $^{12}$CO data,
taking into account the empirical relation between the integrated
emission \hbox{$W_{CO}$,$(=\int T dv$)} and $N_{H2}$. We adopted
$N_{H2}$ = (1.9$\pm$0.3) $\times$ W$_{CO}$ $\times$ 10$^{20}$ cm$^{-2}$
(K \kms)$^{-1}$, obtained from $\gamma$-ray studies of molecular clouds in the Cepheus Flare (Digel \etal1996; Denier \& Lebrun~1990). To derive the molecular mass associated with \hbox{LBN\,148.11--0.45} we took into account the region enclosed by the contour level of \hbox{1.5 K} centered at \hbox{{\it(l,b)} = (148\degr 10\arcmin, --0\degr  20\arcmin)}. For the case of the molecular mass associated with the \hi~shell centered at \hbox{{\it(l,b)} = (149\degr 0\arcmin, --1\degr 30\arcmin)}, we considered the region enclosed by the dash-dot line (Fig.\, 9). Errors in the derived molecular masses arise in uncertainties in distance and in N$_{H_{2}}$.

The ambient density was derived by distributing the ionized, neutral atomic and molecular
 masses over a sphere of radius $R_{e}$. For the case of \hbox{LBN\,148.11--0.45}, $R_{e}$ = 31\arcmin~\hbox{($\equiv$ 9 pc)}, which corresponds to the region showing CO emission. \\
\indent The comparison between the low neutral atomic mass related to \hbox{LBN\,148.11--0.45} and the associated molecular mass suggests that the \hi~mass originated in the photodissociation of the molecular gas. This region is clearly evolving in a highly dense ambient medium.

\section{The origin of the structures}

\subsection{SH\,148.83--0.67}

We can estimate the number of UV photons $N_{Ly-c}$
(s$^{-1}$) necessary to ionized the gas in the \hii\ region
from the radio continuum results and compare our estimate with the UV
photon flux emitted by the exciting star, $N_*$, as derived from stellar
atmosphere models. From the expression for the Stromgren's sphere (see Spitzer 1978), and adopting a radius of 8.3 pc for the \hii~region SH\,148.83--0.67 and the electron
density listed in Table\,4 corresponding to f = 1, we derived
log $N_{Ly-c}$ = 47.9.

Following Smith et al. (2002) and Martins et al.
(2005), the number of Lyman continuum photons emitted by an \hbox{O9 III} star is
in the range $log \ N_*$ = 47.9 - 48.4, indicating that HD\,24431 can
maintain SH\,148.83--0.67 ionized if a small amount of interstellar dust is related to the structure (Inoue 2001).
 Indeed, the lack of infrared emission related to this \hii~region suggests that little interstellar dust is present (see also Table\,5).

\begin{figure*}
\centering
\includegraphics[angle=0,width=0.65\textwidth]{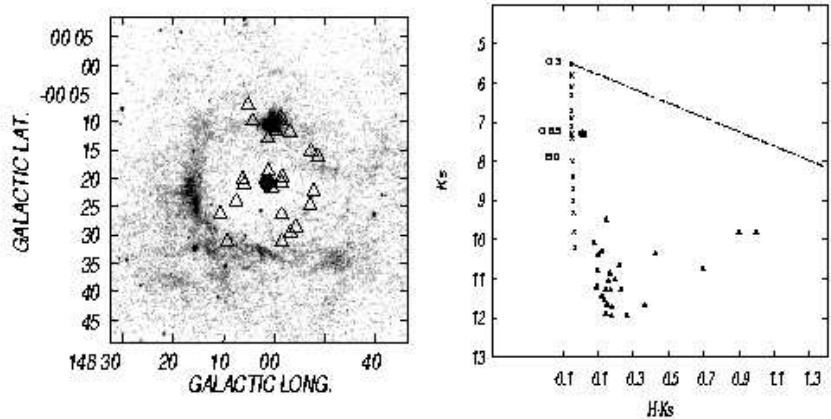}
\caption{\small {\it Left panel}: Spatial distribution of the main sequence star candidates from the 2MASS PSC (triangles) superpossed onto the image at \hbox{8.3$\,\umu$m of LBN\,148.11--0.45}. {\it Right panel}: Near-Infrared Color-Magnitude diagram of the main sequence star candidates shown in left panel. The sources are represented by triangles. The position of the main sequence is shown by crosses. We have adopted a distance of 1.0 kpc. The dashed line represents the reddening curve for an O3 star. The filled circle indicates the position of HD\,24094 in both panels.}
\end{figure*}

To investigate whether the ring-like structure originated in the action of the stellar winds
of HD\,24431 on the surrounding gas, we will estimate the energy conversion efficiency, $\epsilon$, defined as the ratio between the kinetic energy E$_{k}$ in the shell and the mechanical energy supplied by the stellar wind E$_{w}$.

To evaluate the kinetic energy in the expanding shell, \hbox{E$_{k}$ = 0.5 M V$^{2}_{exp}$}, we consider that the ionized, \hi, and molecular gas expand at V$_{exp}$. The kinetic energy results to be \hbox{E$_{k}$ =  2.3 $\times$ 10$^{47}$ erg}.

A rough estimate of the stellar wind energy of HD\,24431, can be obtained adopting a mass loss rate \hbox{\.{ M} =1 $\times$ 10$^{-7}$ M$_{\odot}$ yr$^{-1}$} (Chlebowski \& Garmany 1991) and a terminal velocity  V$_{w}$= 2000 \kms~(Prinja \etal 1990). We assume that this wind blew during at least the dynamical age of the bubble, \hbox{t$_{d}$ = 0.55 R$_{s}$/V$_{exp}$} (McCray 1983). For the values listed in Table\,6,  \hbox{t = 1.1 $\times$ 10$^{6}$ yr}. The lifetime in the main sequence can be considered as an upper limit to this time. From Schaller \etal(1992), \hbox{t = 4.0 $\times$ 10$^{6}$ yr}. For \hbox{t = (1.1 -- 4.0) $\times$ 10$^{6}$ yr}, we estimate \hbox{E$_{w}$ = 0.5 \.M V$^{2}_{w}$ t = (5--16) $\times$ 10$^{48}$ erg}.

The parameter $\epsilon$ = 0.015 -- 0.05, similar to values derived for other interstellar bubbles (Cappa \etal2003), indicating that SH\,148.83--0.67 probably originated in the stellar winds of the O9 III star.

\subsection{LBN\,148.11--0.45}

Adopting the radius and electron density derived for this region from Table\, 4, we estimate that the UV photon flux necessary to ionize the gas is \hbox{log $N_{Ly-c}$ = 47.7}. Clearly this UV photon flux is higher than that provided by HD\,24094 (B8V) and ALS\,7793 (B1V) star.
To look for additional sources of ionizing photons associated with the
\hii\ region, we investigated the presence of main sequence O-type stars
which may contribute to the ionization of the \hii\ region by using the
2MASS data base. We identified the main sequence candidates based on the position of
the sources in the color-magnitude (CCD) and color-color (CMD) diagrams according to the {\it q} parameter. Following Hanson \etal(1997), the parameter {\it q} is defined as:
\begin{equation} \label{factorq}
q= (J-H) - 1.83 \times (H-Ks)
\end{equation}
Main sequence stars have q-values in the range --0.15 to 0.10. Using this selection criteria, we found 28 candidates to be main sequence stars in the region of LBN\,148.11--0.45. Fig.\,10 shows the spatial distribution of these sources superpossed onto the image at 8.3 $\mu$m, and the CMD. Most of the sources have Ks-magnitudes greater than 10 mag, suggesting that they are foreground late-type stars (later than B1 V). HD\,24094, indicated in Fig. 10 as a filled circle, has values (H-Ks, Ks) = (0.023,7.35), corresponding to an O-type star \hbox{(O8.5 or O9)} with A$_{v}$ = 10 mag, different from the B-type star determination by Duerbeck~(1997). Were HD\,24094 an \hbox{O8.5 V --  O9 V} star, it would supply an UV photon flux \hbox{N$_{*}$ = 1 $\times$ 10$ ^{48}$ s$ ^{-1}$} (Martins \etal2001), enough to explain the ionization of the \hii~region. Spectroscopic data are nedded to confirm this point. 

Note that a number of B-type star candidates are projected onto  NVSS\,J035327+533601.

We can roughly estimate the age of the \hii\ region LBN\,148.11--0.45 from the expression   (12-20) by Spitzer (1978), which considers the expansion of an \hii~region with time. The radius of the \hii\ region at the beginning of the expansion phase, $R_i$,  can be derived from the expression for the Strongrem's sphere, adopting an electron density equal to the ambient density (see Table\,6) and \hbox{log $N_{Ly-c}$ = 47.7}. For R$_{i}$ = 0.35 pc and adopting the present radius R$_{i}$ = 7.2 pc, the age of the \hii~region is \hbox{4 $\times$ 10$^{6}$ yr}.

\subsection{Shell centered at (l,b) = (149\degr0\arcmin, --1\degr30\arcmin)}

The question that raises is whether this neutral gas structure is physically related to NGC\,1444. The estimated age of this cluster is in the range \hbox{(25-41)  $\times$ 10$^{6}$ yr} (Lynga 1983, 1987). The spectral type of its earliest star is B2 (Lynga~1983) and four blue stragglers were found to be related to it (Ahumada \& Lapasset~1995). Leisawitz (1990) carried out a molecular survey around open clusters and found molecular clouds within the velocity range [--0.4, --26] \kms~probably associated with NGC\,1444. Clouds labelled by Leisawitz (1990) as I and E centered at (l,b) = (149\degr12\arcmin, --0\degr24\arcmin) and (l,b) = (149\degr36\arcmin, --1\degr6\arcmin) with mean radial velocities of --4.5 and --8.1 \kms, respectively, coincide in position and velocity with the CO ring shown in Fig. 9. No other OB stars projected onto the cavity were found in the literature. The origin of this shell is still an open question.

\section{Summary}

We have examined the interstellar medium in the surroundings of Sh2-205. We based our study on radio continuum data at 408 and 1420 MHz, and 21--cm \hi~line emission data belonging to the Canadian Galactic Plane Survey, $^{12}$CO from the survey by Dame \etal(2001),  IRAS (HIRES) and MSX data.

The main findings can be summed up as follows:

1. Sh2-205 can be separated in three independent structures called: SH\,149.25--0.00, SH\,148.83--0.67, and LBN\,148.11--0.45. To obtain the physical properties of these optical nebulae we have adopted a distance of 1.0$\pm$0.2 kpc.

2. We conclude that SH\,148.83--0.67 is an interstellar bubble powered by the energetic winds of HD\,24431. Low infrared emission  is detected in this region. The \hi~shell related to SH\,148.83--0.67 is identified in the velocity range from --22 to --32 \kms. It is 31 $\times$ 23 pc in size and expands at 6 \kms. The electron density and the associated ionized mass are 15 $\pm$ 4 cm$^{-3}$ and 180 $\pm$ 90 \msun, respectively, adopting a filling factor f = 1. The associated neutral atomic mass is 65 \msun.

3. LBN\,148.11--0.45 is a classical \hii~region. The optical, mid- and far- infrared emissions reveal that dust and gas are well mixed in this region. The dust temperature is $\approx$ 29 K, typical of \hii~regions.  Molecular gas within the velocity range [--0.65, --11.1] \kms~is found to be associated with the \hii~region. We estimated a molecular mass of \hbox{60 $\times$ 10$^{3}$ \msun}. The distribution of the emission in the MSX band A encircles the \hii\ region and correlates with the molecular emission, indicating the presence of a PDR. Neutral atomic gas in the velocity range [+1.6, --18.0] \kms~is related to LBN\,148.11--0.45. \hi~gas with velocities  in the range from +1.6 to --5.2 \kms~is seen projected onto the central part of the \hii~region, which may represent part of the receding cap of the envelope. The \hi\ structure is 15 pc $\times$ 14 pc in size and expands at about 9 \kms. The associated neutral atomic mass is 300 \msun. The electron density and ionized mass are 12 $\pm$ 1 cm$^{-3}$ and 70 $\pm$ 25 \msun, respectively. The \hii~region is evolving in a dense interstellar medium. We estimate that the \hii~region has been expanding for \hbox{4 $\times$ 10$^{6}$ yr}.

The ionizing source of LBN\,148.11--0.45 remains uncertain. The earliest spectral type stars seen in projection onto this region are ALS\,7793 (B1 V) and HD\,24094. Duerbeck (1997) determined a B8 V spectral type for HD\,24094. However, results from the 2MASS point source catalogue showing main sequence stars in the area of the nebula suggest that HD\,24094 is an O8.5 -- O9 star. A star with this spectral type would provide the necessary UV photon flux to keep the \hii~region ionized. High quality spectroscopic data are nedded to elucidate this point.

4. An \hi~structure within the velocity range --5.2 to --15.1 \kms~was identified centered at  \hbox{{\it (l,b)} = (149\degr 0\arcmin, --1\degr 30\arcmin)}. This shell correlates morphologically with molecular gas emission detected in the velocity range from  --0.65 to --11.1 \kms. The neutral and molecular mass are 1600 \msun~and 20 $\times$ 10$^{3}$ \msun, respectively. Possible progenitors of this structure are the stellar members of NGC\,1444.

5. An \hi~shell was also found possibly associated with SH\,149.25--0.00. The neutral atomic structure is expanding at 6 \kms~and is about 17 pc in size. The radio continuum emission is too faint to allow us to determine the physical properties of this nebula. No early type stars were found responsible for the ionization and expansion of \hbox{SH\,149.25--0.00} in the literature.

6. A relatively large number of main sequence candidates were found in the region of \hbox{LBN\,148.11--0.45} by making use of the 2MASS point source catalogue. Particularly interesting are the main sequence candidates projected onto \hbox{NVSS J035327+533601} at \hbox{{\it (l,b)} = (148\degr0\arcmin, --0\degr10\arcmin)}. \\

Finally, the presence of a photodissociation region at the interface between the molecular material and the ionized gas in LBN\,148.11--0.45, strikingly bordering the ionized region, is suggestive of the existence of regions of active stellar formation. This suggestion is reinforced by the high density medium where the \hii~region is evolving. \\
\indent The observed scenario favors the conditions for the {\it collect and collapse} process (Elmegreen 2000), as was found by Deharveng \etal(2003) for Sh2-104. In a forthcoming paper, we analyze the stellar formation activity in the area.

\section*{Acknowledgments}

We thank Dr. T. Dame for making his CO data available to us. We are grateful to the anonymous referee for their helpful suggestions.
This project was partially financed by the
Consejo Nacional de Investigaciones Cient\'{\i}ficas y T\'ecnicas
(CONICET)
of Argentina under project PIP 5886/05, Agencia PICT
14018,
and UNLP under projects 11/G072. The Digitized Sky Survey (DSS) was produced at the Space Telescope Science Institute under US Government grant NAGW-2166.

\label{lastpage}

\end{document}